\documentclass[%
prl,
twocolumn,
superscriptaddress,
 amsmath,amssymb,
 aps,
]{revtex4-2}

\usepackage{booktabs}    
\usepackage{multirow}
\usepackage{graphicx}
\usepackage{dcolumn}
\usepackage{bm}
\usepackage[colorlinks=true,urlcolor=blue,citecolor=blue,linkcolor=blue]{hyperref}
\usepackage[mathscr]{euscript}
\usepackage{array}
\usepackage{siunitx}
\usepackage{soul}

\begin{document}


\title{Clogging of cohesive particles in a two-dimensional hopper}

\author{Johnathan Hoggarth}%
 \thanks{These authors contributed equally to this work. Present address: Mechanical Engineering, Yale University, CT, 06511, USA}
 \affiliation{Department of Physics and Astronomy, McMaster University, 1280 Main Street West, Hamilton, L8S 4M1, Ontario, Canada}

\author{Pablo E. Illing}%
\thanks{These authors contributed equally to this work. Present address: Department of Chemical Engineering, Auburn University, AL, 36849, USA}
\affiliation{Department of Physics, Emory University, Atlanta, Georgia 30322, USA}%

\author{Eric R. Weeks}%
\affiliation{Department of Physics, Emory University, Atlanta, Georgia 30322, USA}%

\author{Kari Dalnoki-Veress}%
\email{dalnoki@mcmaster.ca}
\affiliation{Department of Physics and Astronomy, McMaster University, 1280 Main Street West, Hamilton, L8S 4M1, Ontario, Canada}%
\affiliation{UMR CNRS Gulliver 7083, ESPCI Paris, PSL Research University, Paris, 75005, France}

\date{\today}

\begin{abstract}

We study clogging of cohesive particles in a 2D hopper with experiments and simulations. The system consists of buoyant, monodisperse oil droplets in an aqueous solution, where the droplet size, buoyant force, cohesion, and hopper opening are varied. Stronger cohesion enhances clogging, a trend confirmed in simulations. Balancing buoyant and cohesive forces defines a cohesive length scale that collapses the data onto a master curve. Thus, under strong cohesion, we find that clogging is governed not by particle diameter, but by the cohesive length scale.

\end{abstract}

\maketitle


Granular materials display complex, often counterintuitive, behavior under flow. 
While clogging at narrow constrictions is expected, its underlying statistics and dynamics remain surprisingly rich, with implications for systems ranging from industrial grain handling to landslides and lava flows.
Clogging is controlled by the dimensionless hopper opening–to–particle diameter ratio, $w/d$~\cite{to_jamming_2001, zuriguel_jamming_2003, to_jamming_2005, zuriguel_jamming_2005, janda_jamming_2008, zuriguel_clogging_2014, thomas_FractionClogging_2015, koivisto2017effect, zuriguel_StatisticalMechanicsClogging_2020,vani2024role,hanlan2025cornerstones}. Other than the hopper opening size, there have also been extensive studies of how hopper flow is affected by system properties such as particle shape~\cite{tangOrientationFlowClogging2016, ashourOutflowCloggingShapeanisotropic2017, goldbergCloggingTwodimensionsEffect2018, hafezEffectParticleShape2021}, hopper angle~\cite{lopez-rodriguez_effect_2019, albaraki_how_2014, yu_clogging_2021,vani2024role},  gravity~\cite{dorboloInfluenceGravityDischarge2013, arevaloRoleDrivingForce2014, arevalo_clogging_2016}, friction~\cite{pourninInfluenceFrictionPolydispersityon2007, ashourSiloOutflowSoft2017, alborziMixingParticleSoftness2023a}, and particle softness~\cite{alborziMixingParticleSoftness2023a,berthoDenseBubbleFlow2006, hong_clogging_2017,tao_soft_2021, alborzi_soft_2022,cheng_hopper_2022, alborzi_mixing_2023}. In many cases, granular materials also have cohesive interactions between particles~\cite{sharma2025experimental}, which can significantly impact flow~\cite{faqih_method_2007, szalay_influence_2015, jarray_wet_2019}. For small powders, van der Waals and electrostatics  cause cohesion~\cite{castellanos_relationship_2005} while in larger particles, cohesion can arise due to liquid capillary bridges~\cite{herminghaus__dynamics_2005}. While cohesion is an important parameter, it is typically difficult to  control~\cite{sharma2025experimental}. Recently, Gans \textit{et al.} developed an experimental system to study the impact of cohesion on properties of granular materials~\cite{gans_cohesion-controlled_2020, gans_effect_2021}. Their results indicated that a cohesive length scale that represents the balance between gravity and cohesion was necessary to describe granular flow~\cite{gans_effect_2021}, similar to that proposed by Ono-dit-Biot~\cite{ono-dit-biot_continuum_2020}. Additionally, Zhang \textit{et al.} investigated the clogging of wet granular materials and found that liquid bridges enhanced clogging which could be predicted using an effective aggregate size~\cite{zhang_liquid_2023}. 

\begin{figure}
    \centering
    \includegraphics[width=1\linewidth]{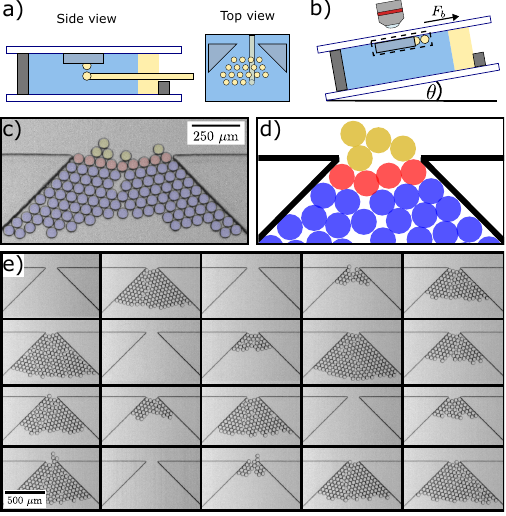}
    \caption{a) Schematic of the experimental chamber. Droplets float into the hopper while the chamber is held horizontal. b) The chamber is rotated to a desired tilt angle which drives the buoyant droplets through the hopper. c) Representative experimental image showing a clog with a false color overlay: red droplets indicate the clogging arch, blue droplets represent remaining droplets, yellow droplets remain attached to the aggregate due to cohesion ($w/d = 8.8$, $w/\delta = 2.86 \pm 0.07$). d) Representative simulation image ($w/d=3.25$, $w/\delta = 3.54$).  e) An array of images of the final state of the hopper for 20 trials with $w/d =$ 3.0, $C_\text{m} = 71$ mM; here 15 out of 20 experiments clogged, so $P_\textrm{clog} = 0.75$.}
    \label{fig:1}
\end{figure}


The ubiquity of inter-particle cohesion motivates a deeper understanding of its role in clogging. Here, we investigate ideal cohesive particles flowing through a 2D hopper (Fig.~\ref{fig:1}). Experimentally, we use monodisperse, buoyant oil droplets in an aqueous solution, varying the droplet diameter $d$, hopper opening $w$, cohesion, and effective gravity. Simulations complement the experimental data, and together they show excellent agreement with a theoretical model.

The experiments are based on previous work by Ono-dit-Biot \textit{et al.} who developed a model system using oil droplets in a surfactant solution to control inter-particle cohesion of a frictionless system~\cite{ono-dit-biot_rearrangement_2020, ono-dit-biot_continuum_2020, hoggarth2025simple}. The surfactant serves two functions: it stabilizes the droplets against coalescence, and, when in excess, forms micelles that induce attractive depletion interactions~\cite{vrij_polymers_1976,bibette_creaming_1992,ono-dit-biot_continuum_2020}. Critically, the surfactant concentration enables precise control of cohesion by tuning the micelle concentration in the solution. Ono-dit-Biot \textit{et al.} proposed a cohesive length scale, which is set by a balance between the inter-particle cohesion and the effect of gravity, similar to that introduced by Gans \textit{et al.}~\cite{gans_effect_2021}, given by $\delta = \sqrt{{\mathscr{A}}/{\Delta\rho g}}$ ~\cite{ono-dit-biot_continuum_2020}. Here $\mathscr{A}=F_c/(\pi R)$ is the cohesive force per unit length, with $F_c$ the unbinding force between two droplets with radius $R$~\cite{brochard2003unbinding}, $\Delta\rho$ is the difference in density between the oil and the aqueous surfactant solution, and $g$ is the acceleration due to gravity. 
In a quasi-2D geometry, where the experimental chamber can be tilted to adjust the effect of gravity (Fig.~\ref{fig:1}), this cohesive length scale is modified to $\delta = \sqrt{{\mathscr{A}}/{\Delta\rho \tilde{g}_e}}$; where we define the effective gravitational acceleration as $\tilde{g_e}= g \sin{\theta}$, with $\theta$ the tilt angle of the chamber~\cite{hoggarth_two-dimensional_2023}.
The 2D system enables direct imaging of the droplets and control over the {effective gravitational force}, $F_b \propto \Delta\rho \tilde{g}$, which drives the particles though the hopper. We note that by `effective gravitational force', we mean the sum of the gravity and buoyancy force acting parallel to the top surface of the chamber.
Previously, the cohesive length-scale, $\delta$,  has been used to describe the spreading of oil droplets in both 2D and 3D~\cite{ono-dit-biot_continuum_2020, hoggarth_two-dimensional_2023}.


Experimentally, a quasi-2D hopper is made by coating a glass slide with a $\sim$100 $\mu$m layer of SU-8 photoresist (Kayaku, USA) and photolithography is used to pattern a hopper with walls at 45$^\circ$ angles. For the experiments, the hopper opening spanned $w \in [100, 500] \,\si{\micro\meter}$, with most experiments performed at 150 $\mu$m and 200 $\mu$m. Experimental chambers are constructed by placing a 30 x 20 x 5 mm$^3$ 3D-printed spacer between the 2D hopper slide and a glass microscope slide and imaged with a camera. A schematic diagram of the experiment is shown in Fig.~\ref{fig:1}(a) and (b). The setup is placed on a rotation stage such that the chamber can be tilted to an angle, $\theta$.

Chambers are filled with an aqueous solution of the surfactant, sodium dodecyl sulfate (SDS), and sodium chloride (NaCl) to screen ionic interactions. The concentration of SDS varies from 7 mM to 265 mM while the concentration of NaCl is held constant at 1.5\% (w/w). 
SDS micelles generate an attractive interaction between the oil droplets, increasing linearly with micelle concentration, $\mathscr{A} \propto C_\text{m}$ where $C_\text{m}$ is the concentration of micelles~\cite{vrij_polymers_1976,bibette_creaming_1992,ono-dit-biot_continuum_2020}. 
Glass capillary tubes (OD 1 mm, ID 0.58 mm, World Precision, USA) are pulled using a pipette puller (Narishige, Japan) to an opening diameter of $\sim 10 \ \mu$m. The pipette 
is inserted into the chamber and light paraffin oil (Supelco, MilliporeSigma, USA) is dispensed to create near-monodisperse droplets with $ d \in [52, 67]\,\si{\micro\meter}$ using the snap-off instability (the coefficient of variation in droplet radius is $\sim 0.5$\%.)~\cite{barkley_snap-off_2015, barkley_predicting_2016}. The buoyant droplets rise and fill the hopper, and because of their small size and correspondingly high Laplace pressure, they can be approximated as hard spheres. The chamber is initially held horizontal and droplets are deposited in a loosely packed pattern; see \href{SupplementalVideo-1.mp4}{supplemental video 1} for the filling process. 

After $190 \pm 5$ droplets have been deposited, the pipette is removed away from the hopper and the chamber is rotated to a desired tilt angle, which initiates the flow of droplets through the hopper. Droplets are monitored over time for the presence of clogs [see Fig.~\ref{fig:1}(c)]. Once all the the droplets flow through the hopper, or if a clog is detected, the experiment is complete and the chamber is rotated past 90$^\circ$ to clear the chamber of the droplets. Once the hopper is emptied, the chamber is rotated back to horizontal, and the experiment is automated to repeat 20 times for a specific set of parameters [see Fig.~\ref{fig:1}(e)]. The probability of clogging, $P_\text{c}$, was measured by dividing the number of clogs by the number of total trials (see \href{SupplementalVideo-2.m4v}{supplemental video 2} and \href{SupplementalVideo-3.m4v}{supplemental video 3}).

For the simulations of the hopper we use a quasi 2D Durian bubble model~\cite{durian_foam_1995}, as modified in Refs.~\cite{hong_clogging_2017,illing_compression_2024}.  This model considers soft particles with large viscous forces acting on them, such that all other forces balance the velocity-dependent viscous drag force.  Thus at each time step the equations of motion are solved for the velocity rather than the acceleration. For each particle $i$ the equation to solve is: $\sum_j\left[  \vec{F}_{ij}^{\mathrm{cont}} + \vec{F}_{ij}^{\mathrm{visc}}  \right] + \vec{F}_i^{\mathrm{wall}} + \vec{F}_i^{\mathrm{grav}}+\vec{F}_i^{\mathrm{drag}}=0$, 
where $\vec{F}_{ij}^{\mathrm{cont}}$ is the contact force between droplets $i$ and $j$, $\vec{F}_{ij}^{\mathrm{visc}}$ is the viscous interaction between two contacting droplets, $\vec{F}_{i}^{\mathrm{wall}}$ is all forces due to interactions with walls (attractive, repulsive and viscous forces~\cite{hong_clogging_2017,illing_compression_2024}), $\vec{F}_i^{\mathrm{grav}}$ is the effective gravitational force, and $\vec{F}_i^{\mathrm{drag}}$ is the viscous drag.  Each droplet is modeled as a sphere:  $\vec{F}_{ij}^{\mathrm{cont}}$ has a spring-like repulsive force for spheres that overlap, along with an attractive depletion force based on the Asakura-Oosawa model~\cite{asakura_Interaction_1954} and the depletion free energy for overlapping spheres from J.C. Crocker et al.~\cite{crocker_Entropic_1999}.  The depletion force is tuned with the parameter $\phi_c$, the effective volume fraction of the micelles, which is a nondimensional analogue of $C_\text{m}$ in the experiments~\cite{hong_clogging_2017,illing_compression_2024}.

The time step in simulations is defined by $t_0=B R /f_0$, the timescale for two droplets to push apart, limited by inter-droplet viscous interactions; where $B=1$ is the viscous coefficient, $R= 1$ is the particle radius,
and $f_0=10$ is the spring constant of a droplet.  With the current parameter choices, the free-fall velocity of an isolated droplet is proportional to $\vec{F}_i^{\mathrm{grav}}/\vec{F}_i^{\mathrm{drag}}$, and so also to the force of gravity; we use the nondimensional acceleration due to gravity in the simulations, $\tilde{g}_s$, as a control parameter for the driving force. For further discussion on the choice of parameters see Refs.~\cite{hong_clogging_2017,illing_compression_2024}.

As in the experiment, 190 particles are simulated with a hopper  angle of $45^{\circ}$, and a droplet polydispersity set at $0.01$. The simulation begins with particles randomly placed in the hopper. They free-fall, settle, and once static, the hopper exit is opened. The force equation can be rewritten as a first order differential equation for the velocity, which is solved using a fourth-order Runge-Kutta algorithm and a time step of $dt=0.1 t_0$. Simulations end when all droplets have exited the hopper, or when the maximum speed of all particles in the hopper is below $10^{-12}$, which defines a clog. For these simulations the choice of parameters for depletion strength and gravity were made based on previous simulations involving cohesive forces~\cite{illing_compression_2024}, and hopper simulations~\cite{hong_clogging_2017}, with the overall goal of maintaining a range of parameters similar to the experiments. 
In the simulations, $\delta=\sqrt{F_c/(\rho \tilde{g}_s \pi  R )  }$, where $F_c$ is the force needed to pull apart two particles from the equilibrium position and $\rho=\frac{3}{4\pi}$~\cite{illing_compression_2024}.
One set of simulations keeps $w/ d  = 3$, and varies $\tilde{g}_s$ and $\phi_c$. A second set of simulations keeps $\tilde{g}_s=0.01$, while varying  $w$ and  $\phi_c$.

\begin{figure}
    \centering
    \includegraphics[width=\linewidth]{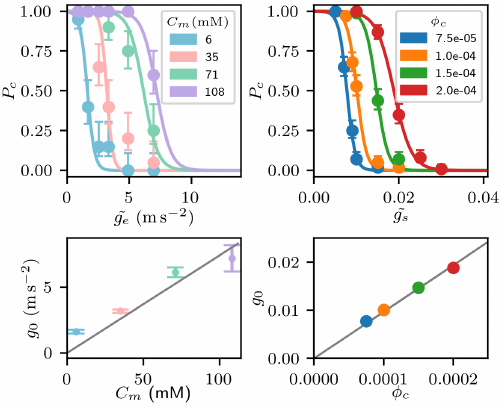}
    \caption{Clogging probability of a) experiments with $w= 162~\mu$m and $d=56~\mu$m and b) simulations with $w/d = 3.0$, for a range of cohesive strengths as a function of the effective gravitational acceleration (error is calculated based on a finite number of trials $N = 20$ and $N = 100$ for a Poisson process). Solid lines are fits of Eq.~(\ref{eqn:P_clog}) to the data. The effective gravitational acceleration where the probability of clogging is $1/2$, $g_0$, vs cohesive strength for c) experiment and d) simulation.  }
    \label{fig:2}
\end{figure}

We first investigate the impact of gravity and cohesion while keeping the width to droplet diameter constant. 
In experiments, gravity is tuned by the hopper tilt angle $\theta$ ($\tilde{g}_e = g\sin\theta$) and cohesion by the micelle concentration $C_\text{m}$. In simulations, $\tilde{g}_s$ and $\phi_c$ control gravity and cohesion.
In Fig.~\ref{fig:2}(a,b) we show $P_\text{c}$ as a function of the effective gravitational acceleration for constant $w/d$. For a constant cohesion strength, as the effective gravitational force increases, the clogging probability decreases. In addition, increasing cohesion increases the clogging probability. These results are physically intuitive: increasing the effect of gravity or decreasing cohesion can destabilize a potential arch, thus decreasing the potential to clog. 

We use a general sigmoidal function to fit the data:
\begin{equation}
   P_\text{c}= \left[1+\exp[(\tilde{g}_i-g_{0})/b]\right]^{-1},
    \label{eqn:P_clog}
\end{equation}
where $b$ is the width of the transition, $g_0$ is the value of the effective gravitational acceleration that results in $P_\text{c}=1/2$, and the subscript \( i \in \{e, s\} \) designates the experimental and simulated values (see Supplemental Table 1 for fit parameters). We plot $g_0$ as a function of $C_\text{m}$ and $\phi_c$ in Fig.~\ref{fig:2}(c,d). 
Both experiment and simulation show a monotonic increase, consistent with the expectation that maintaining a constant clogging probability ($P_\text{c} = 1/2$) requires a greater driving force with increasing cohesion.

Having investigated the role of the driving force and cohesive strength in Fig.~\ref{fig:2}, we now turn to the impact of the hopper opening size $w/d$.  We will show how $w/d$ relates to the clogging probablity $P_\text{c}$ and the mean avalanche size, $\langle s \rangle$, which is defined as the number of particles passing through the hopper before a clog. Theoretical and empirical relations related to the avalanche size and probability of clogging have been well established~\cite{to_jamming_2001, zuriguel_jamming_2003, to_jamming_2005, zuriguel_jamming_2005, janda_jamming_2008, zuriguel_clogging_2014, thomas_FractionClogging_2015, koivisto2017effect, zuriguel_StatisticalMechanicsClogging_2020}. Here, the following is inspired by the work of Janda \textit{et al.}~\cite{janda_jamming_2008} and the derivations by Durian \textit{et al.}~\cite{thomas_FractionClogging_2015,koivisto2017effect}; we recap the essential physics and extend to the case of cohesive particles. If $\langle s \rangle$ is the number of  particles in an avalanche, then $\langle s \rangle+1$ particles cause a clog, and the probability of any particle causing a clog is given by $p_\text{c}=(\langle s \rangle+1)^{-1}$.  We can then write $\langle s \rangle = (1-p_\text{c})/p_\text{c}$, which is the ratio of the probability that a particle passes through the orifice, $1-p_\text{c}$, to the probability the particle causes a clog, $p_\text{c}$. A clog is formed by an arch in 2D or dome in 3D. Assuming that each grain within the arch or dome can exist in $\omega$ microstates, of which $\omega_c$ are states that can result in a clog, then the probability that a particle is in a clogging state is $\omega_c/\omega$. Now, it is not enough for a single particle to be in a clogging state, all particles that form an arch or dome must also be. We can then write $p_\text{c}=\kappa(\omega_c/\omega)^n$, where $n$ is the number of particles in the arch or dome, and $\kappa$ is a normalization constant. 

We now turn to the number of particles, $n$, that form an arch or dome, which must depend on the size of the opening. It has been suggested and empirically  verified~\cite{to_jamming_2005,janda_jamming_2008,thomas_FractionClogging_2015, zuriguelSiloCloggingReduction2011} that $n\propto (w/d)^\alpha$, where $\alpha=2$ for an arch, and $\alpha=3$ for a dome. First of all, we stress that while this scaling is valid for cohesionless particles, it must be modified when cohesion is important, as we will do below. Second, we wish to clarify the dependence on the dimensionality of the system, because it is often assumed that an arch is a string of particles and a dome is a sheet of particles (i.e. $\alpha =1$ for an arch and 2 for a dome). An arch or dome alone must be stabilized by neighboring particles (i.e. a 1D arch of marbles is not stable). Thus, these structures have some thickness to them, and the natural scale for this is the width of the opening. There are two parts to this argument: 1) an arch or dome must have a “thickness” to it; and 2) that thickness must scale with the width of the opening, resulting in a self-similar clog structure. For the 2D case studied here, $\alpha =2$, and $n={\beta(w/d)^2}$,
with geometric prefactor $\beta$. Then  
$\langle s \rangle = \left[\kappa\left(\omega_c/\omega\right)^{\beta(w/d)^\alpha}\right]^{-1}-1$.
The constant $\kappa=(\omega_c/\omega)^{-\beta}$, which follows from $\langle s \rangle=0$ at $w/d=1$. Then,
\begin{equation}
\langle s \rangle = \exp\left\{ {\cal C} \left[ ( w/d)^\alpha - 1 \right] \right\} - 1,
\label{sfinal}
\end{equation}
with constant ${\cal C}$. This expression is properly normalized and of a similar form to that used by others~\cite{janda_jamming_2008,thomas_FractionClogging_2015,koivisto2017effect}.

The probability of some particular particle causing a clog is $(\langle s \rangle +1)^{-1}$; thus the probability of $N$ particles in the hopper not clogging is $[1-(\langle s \rangle +1)^{-1}]^N$.  Conversely, the probability of clogging with $N$ particles is $P_\text{c}(N)=1-[1-(\langle s \rangle +1)^{-1}]^N=1-\exp[-N \ln (1+\langle s\rangle^{-1})]$. Making the approximation that $\ln (1+\langle s\rangle^{-1}) \approx  \langle s\rangle^{-1}$, we can write the probability of clogging with $N$ particles as:
\begin{equation}
P_\text{c}(N) \approx 1-\exp \left( -N/ \langle s\rangle \right),
\label{P_c}
\end{equation}
where $\langle s \rangle$ is given by Eq.~(\ref{sfinal}).  Thus, we have related $\langle s \rangle$ and $P_\text{c}$ to $w/d$ via Eqs.~(\ref{sfinal}) and (\ref{P_c}) with $\alpha=2$ for our 2D system, $N=190$ for our number of droplets, and ${\cal C}$ as the sole fitting parameter.

\begin{figure}
    \centering
    \includegraphics[width=\linewidth]{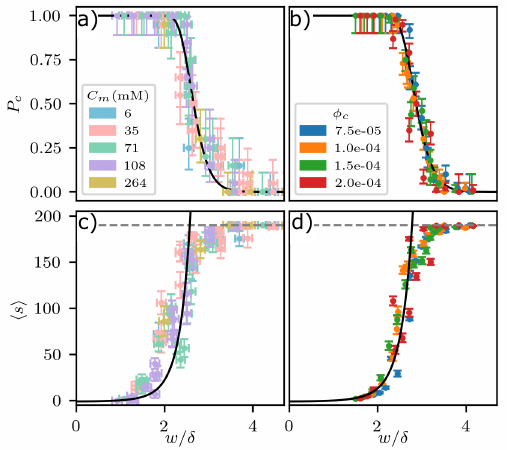}
    \caption{Probability of clogging as a function of $w/\delta$ for (a) experiments and (b) simulations. The solid lines represents a fit to Eq.~(\ref{P_c}). 
    (c) Mean avalanche size before a clog occurs as a function of $w/\delta$ for experiments and (d) simulations. Solid lines are Eq.~(\ref{sfinal}) fit to the data. Dashed lines show a plateau at 190 droplets. Error bars on $P_c$ are calculated based on a finite number of trials $N = 20$ for experiments and $N = 100$ for simulations for a Poisson process. Error bars on $\langle s \rangle$ represent standard error on the mean. Error bars on $\delta$ are representative of experimental data measuring the cohesive strength between droplets~\cite{hoggarth2025simple}.  For these data, smaller $w/\delta$ represents a smaller opening or stickier droplets, both of which make clogging likelier.}
    \label{fig:3}
\end{figure}

This discussion has neglected cohesion.  For strong inter-particle cohesion, we propose that the critical length scale that determines clogging is not the droplet diameter $d$, but rather the cohesive length scale $\delta$. This assumption is consistent with observations made by Gans \textit{et al.}~\cite{gans_effect_2021} and other works~\cite{ono-dit-biot_rearrangement_2020, ono-dit-biot_continuum_2020, hoggarth_two-dimensional_2023, illing_compression_2024}. 
Conceptually, with increasing cohesion, the relevant size is that of cohesively stabilized aggregates of size $\delta$, not individual droplets. This effectively renormalizes $w/d$ to $w/\delta$ in Eq.~(\ref{sfinal}). In our experiments, $ d\in [52, 67]\,\si{\micro\meter}$, while $\delta \in [30.5 \pm 0.3, 260 \pm 10]\, \si{\micro\meter}$~\cite{ono-dit-biot_continuum_2020}. At low cohesion, clogging depends on $d$, but our focus is on regimes where $\delta$ dominates.
The probability of clogging, $P_\text{c}$, as a function of $w/\delta$ is shown in Fig.~\ref{fig:3}(a,b) for 91 experiments each consisting of 20 trials (5 droplet radii and 5 cohesion strengths) and for 51 simulations each consisting of 100 trials ($w \in [1.75d, 6d]$ and 4 cohesion strengths). 
Remarkably, all data collapse onto a single master curve which is well fit by Eq.~(\ref{P_c}) with ${\cal C}=0.76\pm0.01$ (experiment) and ${\cal C}=0.690\pm0.004$ (simulation), confirming that the dimensionless length scale $w/\delta$ captures the clogging behavior of cohesive particles.  Given that $\delta > d$ for the cases we are considering, clogging occurs for larger opening sizes $w$ than could clog for cohesionless particles.

To further investigate the importance of $w/\delta$ to our system, plots of $\langle s \rangle$ as a function of $w/\delta$ are shown in Fig.~\ref{fig:3}(c,d). Again, the data collapse onto a single master curve. Since the number of particles in the hopper was limited to $N\sim 190$, the plots in Fig.~\ref{fig:3}(c,d) plateau at $\langle s \rangle \sim N$. The plateau does not imply that we have reached a critical value for which clogging never occurs, rather we have reached a region where the probability of clogging for 190 particles vanishes. We compare Eq.~(\ref{sfinal}) to the data of Fig.~\ref{fig:3}(c,d) \textit{with the same} fit values for ${\cal C}$ used in Fig.~\ref{fig:3}(a,b). The ability of the data to collapse onto a master curve as well as the goodness of fit to Eq.~(\ref{sfinal}) further confirms that the critical parameter describing clogging is given by $w/\delta$. This also validates the assumptions made in the derivation of Eqs.~(\ref{sfinal}) and (\ref{P_c}):  in particular, the number of particles stabilizing an arch (in a two-dimensional system) scales as $w^2$, as seen in prior work with cohesionless particles \cite{to_jamming_2005,janda_jamming_2008,thomas_FractionClogging_2015, zuriguelSiloCloggingReduction2011}.

Lastly, we return to the empirical trends shown in Fig.~\ref{fig:2}(c,d) for the relationship between the driving force, $F_b\propto g_0$, and cohesion strength for which $P_\text{c}=1/2$. We see from Eq.~(\ref{P_c}) that $P_\text{c}=1/2$ corresponds to $w/\delta$ being some constant: when $P_\text{c}=1/2$, $w \propto \delta$. Since $\delta =  \sqrt{{\mathscr{A}}/{\Delta\rho \tilde{g}}}$, we obtain $\tilde{g}w^2\propto \mathscr{A}$. With Eq.~(\ref{eqn:P_clog}) and setting $P_\text{c}=1/2$, we have $g_0=\tilde{g}$. 
We then obtain $g_0 w^2 \propto \mathscr{A}$, where $\mathscr{A}$ is proportional to $C_\text{m}$ (experiment) and $\phi_c$ (simulation). The data in Fig.~\ref{fig:2}(c,d), where the opening width is fixed, follow this linear relationship closely, providing independent confirmation that $\delta$ is the relevant length scale governing clogging for cohesive particles.

In conclusion, we have observed the clogging of frictionless cohesive particles in a 2D hopper and outlined the impact of the cohesive strength on the ability of the particles to flow. We find a clear dependence of clogging on the cohesive strength of the particles. Furthermore, we have demonstrated that a fundamental cohesive length scale $\delta$ is critical to describing clogging, which can be obtained by balancing the cohesive strength with the effective gravitational force. When plotting both the mean avalanche size and the clogging probability as a function of a dimensionless ratio $w/\delta$, our data collapse on master curves (Fig.~\ref{fig:3}).  The collapse matches theory using $\delta$ as the key length scale, underscoring the importance of cohesion.  Remarkably, this allows clogging to occur even with large hopper openings such as in the experiments shown in Fig.~\ref{fig:1}(c), for which clogging would not be possible for a cohesionless system \cite{hong_clogging_2017}.
Under strong cohesion, clogging is governed not by particle diameter but by a cohesive length scale.


\begin{acknowledgments}
JH and K.D-V. acknowledge financial support from the Natural Science and Engineering Research Council of Canada. JH acknowledges funding from the  Canada Graduate Scholarship. This material is based upon work supported by
the National Science Foundation under Grant No. CBET-2306371. The authors are grateful for valuable discussions with Douglas Durian and \'Elie Rapha\"el.
\end{acknowledgments}


\vfill\eject

\onecolumngrid

\section{Supplemental Figures and tables}

\begin{figure}[h!]
    \centering
    \includegraphics[width=0.6\linewidth]{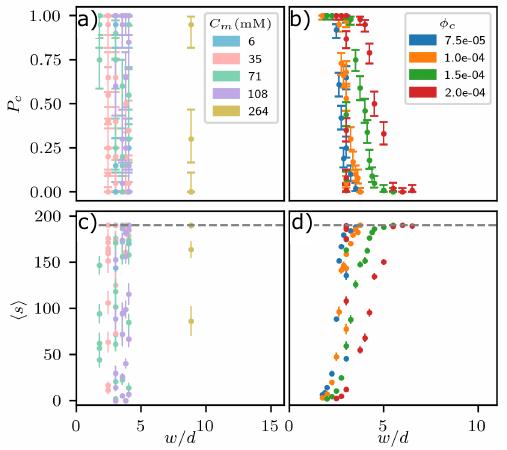}
    \caption{Probability of clogging as a function of $w/d$ for (a) experiments and (b) simulations. (c) Mean avalanche size before a clog occurs as a function of $w/d$ for experiments and (d) simulations. Dashed lines show a plateau at 190 droplets. Error bars represent standard error on the mean.  For these data, we see that simply plotting as a function of $w/d$ does not nicely collapse our data which contains different cohesion strengths. In short, $d$ cannot normalize the data, while $\delta$  does. }
    \label{fig:supplemental}
\end{figure}


\begin{table}[h]
    \centering
    \caption{
        Fit parameters \(g_{0}\) and \(b\) for Eq.~(1) in the main text:
        \(P_\mathrm{c} = \left[1+\exp\left((\tilde{g}_i - g_{0})/b\right)\right]^{-1}\).
        Left: experimental results for concentrations \(C_m\).
        Right: simulation results for depletion strength \(\phi_c\).
        Uncertainties represent one standard deviation. 
        For the $C_m = 108$ mM experimental data, there is a limited number of data points. We could not go to higher values of the tilt angle (larger $g_e$) to obtain more data because then the tilt angle is so large that 3D packing of the droplets occurs. However within the 2D packing regime fortunately there is one non-trivial point which is close to $Pc \sim 0.5$ and can give a reliable measure of $g_0$. While this leads to an increased uncertainty in that $g_0$ data point, the value is in good agreement with the linear trend established by the other experiments.  Note also that in Figs. 2(c,d), the linear fit is required to go through the origin:  when the cohesive forces are zero, the opening width is wide enough that no clogging occurs.  This requirement is another constraint on the fit.
    }
    \setlength{\tabcolsep}{10pt}
    \begin{tabular}{lcc|lcc}
        \toprule
        \multicolumn{3}{c|}{\textbf{Experiment}} & \multicolumn{3}{c}{\textbf{Simulation}} \\
        \midrule
        \(C_m  \textrm{ (mM)}\) & \(g_{0} \textrm{ (m s}^{-2})  \) & \(b  \textrm{ (m s}^{-2}) \) & \(\phi_c\) & \(g_{0}\) & \(b\) \\
        \midrule
        6   & \(1.7 \pm 0.2\)  & \(0.4 \pm 0.1\)  & 7.5$\times 10^{-5}$ & \((8.0 \pm 0.2)\times 10^{-3}\)  & \((0.75 \pm 0.07)\times 10^{-3}\) \\
        35  & \(3.1 \pm 0.3\)  & \(0.5 \pm 0.2\)  & 1$\times 10^{-4}$   & \((10.1 \pm 0.2)\times 10^{-3}\) & \((0.9 \pm 0.2)\times 10^{-3}\)  \\
        75  & \(5.9 \pm 0.5\)  & \(0.9 \pm 0.3\)& 1.5$\times 10^{-4}$ & \((14.7 \pm 0.3)\times 10^{-3}\) & \((1.1 \pm 0.2)\times 10^{-3}\)  \\
         108& \(7.2 \pm 1.0\) & -- & 2$\times 10^{-4}$   & \((19.0 \pm 0.4)\times 10^{-3}\) & \((1.9 \pm 0.3)\times 10^{-3}\)  \\
        \bottomrule
    \end{tabular}
    \label{tab:fits}
\end{table}

\begin{table}[h]
    \centering
    \caption{
        Values of \(\tilde{g}_e\) (experiment) and \(\tilde{g}_s\) (simulation) 
        with corresponding values of \(\delta\).  
        Left: experimental results for concentrations \(C_m\).  
        Right: simulation results for depletion strength \(\phi_c\).
    }
    \setlength{\tabcolsep}{8pt}
    \begin{tabular}{lcc|lcc}
        \toprule
        \multicolumn{3}{c|}{\textbf{Experiment}} & \multicolumn{3}{c}{\textbf{Simulation}} \\
        \midrule
        \(C_m \, (\mathrm{mM})\) & \(\tilde{g}_e  \textrm{ (m s}^{-2})  \) & \(\delta\ (\mu\textrm{m})\)  
        & \(\phi_c\) & \(\tilde{g}_s\) & \(\delta\) \\
        \midrule
        \multirow{6}{*}{6} 
            & 0.88 & 86.9 & \multirow{5}{*}{7.5$\times 10^{-5}$}& 0.015
& 0.733\\
            & 1.67 & 61.5 & & 0.01
& 0.898\\
            & 2.55 & $50.4$ & & 0.009
& 0.946\\
            & 3.33 & $43.8$& & 0.007
& 1.07\\
            & 4.9 & $36.3$& & 0.005
& 1.27\\
            & 6.96 & $30.5$& & & \\
        \midrule
        \multirow{6}{*}{35} 
            & 0.88 & 120& \multirow{6}{*}{1$\times 10^{-4}$}& 0.03
& 0.603\\
            & 1.67 & 84.3& & 0.02
& 0.739\\
            & 2.55 & 69.0& & 0.015
& 0.853\\
            & 3.33 & 60.0& & 0.01
& 1.04\\
            & 4.9 & 49.7& & 0.009
& 1.10\\
            & 6.96 & 41.8& & 0.0075
& 1.21\\
        \midrule
        \multirow{6}{*}{71}& 0.88 & 150& \multirow{3}{*}{1.5$\times 10^{-4}$}& 0.02
& 0.920\\
            & 1.67 & 106& & 0.015
& 1.06\\
            & 2.55 & 86.7& & 0.01
& 1.30\\
            & 3.33 & 75.5& & & \\
            & 4.9 & 62.4& & & \\
            & 6.96 & 52.5& & & \\
        \midrule
        \multirow{6}{*}{108} 
            & 0.88 & 175& \multirow{5}{*}{2$\times 10^{-4}$}& 0.03
& 0.882\\
            & 1.67 & 124& & 0.025
& 0.967\\
            & 2.55 & 102& & 0.02
& 1.08\\
            & 3.33 & 88.5& & 0.015
& 1.25\\
            & 4.9 & 73.2& & 0.01
& 1.53\\
            & 6.96 & 61.6& & & \\
        \bottomrule
    \end{tabular}
    \label{tab:gtilde_delta}
\end{table}

\begin{table}[h]
    \centering
    \caption{
        Fit parameter \(C\) for Eq.~(5) in the main text: 
        \(\langle s \rangle = \exp\left\{ C \left[ \left( \frac{w}{d} \right)^\alpha - 1 \right] \right\} - 1\).
        Left: experimental results for concentrations \(C_m\).
        Right: simulation results for depletion strength \(\phi_c\).
        Uncertainties represent one standard deviation.
    }
    \setlength{\tabcolsep}{12pt}
    \begin{tabular}{l c | l c}
        \toprule
        \multicolumn{2}{c|}{\textbf{Experiment}} & \multicolumn{2}{c}{\textbf{Simulation}} \\
        \midrule
        \(C_m \textrm{ (mM)}\) & \(C\) & \(\phi_c\) & \(C\) \\
        \midrule
        6   & 1.23 $\pm$ 0.08      & 7.5$\times 10^{-5}$ & (7.7 $\pm$ 0.1)$\times 10^{-3}$ \\
        35  & 3.21 $\pm$ 0.15      & 1$\times 10^{-4}$   & (10.1 $\pm$ 0.2)$\times 10^{-3}$ \\
        75  & 5.94 $\pm$ 0.12      & 1.5$\times 10^{-4}$ & (14.7 $\pm$ 0.3)$\times 10^{-3}$ \\
        108 & 7.30 $\pm$ 0.10      & 2$\times 10^{-4}$   & (18.8 $\pm$ 0.2)$\times 10^{-3}$ \\
        \bottomrule
    \end{tabular}
    \label{tab:C_fits}
\end{table}

\end{document}